\documentstyle[prl,aps,epsf]{revtex}

\begin{document}
\def \beq {\begin{equation}}
\def \eeq {\end{equation}}
\def \bes {\begin{eqnarray}}
\def \ees {\end{eqnarray}}
\def \ni {\noindent}
\def \nn {\nonumber}
\def \z {\tilde{z}}
\def \kp {k_{\bot}}
\def \e {\varepsilon}
\def \dpa {\Delta_{\|}}
\def \dpp {\Delta_{\bot}}

\title{Comment on ``Does the transverse electric zero mode
contribute to the Casimir effect for a metal?
}

\author{
V.~B.~Bezerra,
G.~L.~Klimchitskaya,\footnote{On leave from
North-West Technical University,
St.Petersburg, Russia.
E-mail: galina@fisica.ufpb.br}
and V.~M.~Mostepanenko\footnote{On leave from
Noncommercial Partnership ``Scientific Instruments'',
Moscow, Russia.
E-mail: mostep@fisica.ufpb.br}
}

\address{Departamento de F\'{\i}sica, Universidade
Federal da Para\'{\i}ba,
C.P.~5008, CEP 58059-970,
Jo\~{a}o Pessoa, Pb-Brazil
}
\maketitle

{\abstract{Recently
J.~S.~H{\o}ye, I.~Brevik, J.~B.~Aarseth, and K.~A.~Milton
[Phys. Rev. E {\bf 67}, 056116 (2003)] proposed that if
the Lifshitz formula is combined with the Drude model, the
transverse electric zero mode does not contribute to the
result for real metals and there arises a linear temperature 
correction to the Casimir force at small temperatures.
The authors claim that in spite of the fact that the Casimir
entropy in their approach is negative, the Nernst heat theorem
is satisfied. In the present Comment we show that 
the authors' conclusion regarding the  Nernst heat theorem 
is in error.
We demonstrate also the resolution of this
thermodynamic puzzle based on the use of the surface impedance
instead of the Drude dielectric function. The results of
numerical computations obtained by the authors are compared with 
those from use of the surface impedance approach
which are thermodynamically consistent.
}}

\vspace{2mm}
PACS number(s): 05.30.-d, 11.10.Wx, 73.61.At, 77.22.Ch

\vspace{3mm}
\large

A recent calculation shows that the transverse electric zero
mode does not contribute to the Casimir free energy \cite{1}.
This results in an observable linear temperature correction
to the Casimir force between parallel plates made of real 
metal \cite{1}. According to Ref.~\cite{1}, the Casimir free
energy increases with increasing temperature in some temperature
interval, and the Casimir entropy becomes negative in this
interval. From the view-point of the authors of Ref.~\cite{1},
this, however, does not lead to violation of the Nernst heat
theorem. Below we show that on the contrary the transverse electric
zero mode does contribute to the Casimir free energy which
results in the absence of a linear in temperature correction
to the Casimir force at small temperatures. In reality,
the Casimir free energy is universally a decreasing function of
temperature, and the Casimir entropy is always positive.
We demonstrate also that if, as is done in Ref.~\cite{1}, the
Drude model and the Bloch-Gr\"{u}neisen formula are used to
calculate the Casimir force, the Nernst heat theorem is
nesessarily violated.

The correct description of the thermal Casimir force
between real metals has assumed great importance due to 
the recent precision experiments \cite{2,3,4,5,6,7,8},
followed by the prospective applications of the Casimir effect
in nanotechnology \cite{9,10} and also its use as a test for predictions
of fundamental physical theories \cite{11,12,13,14}. Unexpectadly,
it was found that the calculations of the thermal Casimir force
on the basis of the Lifshitz formula \cite{15} supplemented by
the Drude model ran into serious difficulties 
\cite{16,17,18,19,20,21,22,23,24,25,26,27,28,29}.
The key question of the controversy is whether the transverse electric
zero mode contributes to the Casimir effect in the case of real
metals. Bostr\"{o}m and Sernelius \cite{16} have used the Drude model
to compute the contributions of both transverse electric and transverse
magnetic modes and found that the transverse electric zero mode
does not contribute. Later on it was shown \cite{26,28} that in
this approach the Casimir entropy is negative in some temperature
interval and the third law of thermodynamics (the Nernst heat
theorem) is violated. On the contrary, the authors of Ref.~\cite{1}
claim that the entropy is zero at zero temperature, i.e. the third law
of thermodynamics is not violated.

To prove this statement they present in Sec.~IV of Ref.~\cite{1} two
sets of arguments, analytical and numerical. Both of them start
from Eq.~(3.4) representing the Lifshitz result for the Casimir free
energy per unit area in the configuration of two semispaces separated
by a distance $a$:
\beq
\beta F=\frac{1}{2\pi}\sum\limits_{m=0}^{\infty}{\vphantom{\sum}}^{\prime}
\int_{\zeta_m}^{\infty}\left[\ln\left(1-\lambda_{m}^{TM}\right)
+\ln\left(1-\lambda_{m}^{TE}\right)\right]q\,dq,
\label{1}
\eeq
\ni
where
\bes 
&&
\lambda_{m}^{TM}=A_me^{-2qa}, \qquad
\lambda_{m}^{TE}=B_me^{-2qa},
\label{2} \\
&&
A_m=\left[\frac{{\e}_mq-\sqrt{({\e}_m-1)\zeta_m^2+q^2}}{{\e}_mq+
\sqrt{({\e}_m-1)\zeta_m^2+q^2}}\right]^2,
\quad
B_m=\left[\frac{q-\sqrt{({\e}_m-1)\zeta_m^2+q^2}}{q+
\sqrt{({\e}_m-1)\zeta_m^2+q^2}}\right]^2.
\nn
\ees
\ni
Here the dielectric permittivity ${\e}_m\equiv\e(i\zeta_m)$
is computed on the imaginary Matsubara frequencies,
$\zeta_m=2\pi m/\beta$, $\beta\equiv 1/T$, and 
units with $k_B=\hbar=c=1$ are used.

According to the Drude model
\beq
\e(i\zeta)=1+\frac{\omega_p^2}{\zeta(\zeta+\nu)},
\label{3}
\eeq
\ni
where $\omega_p$ is the plasma frequency, $\nu=\nu(T)$ is the
relaxation parameter. Substituting Eq.~(\ref{3}) into Eq.~(\ref{2})
one obtains $B_0(0)=0$, i.e. the transverse electric zero mode
does not contribute in the case of the Drude metal in accordance
with Refs.~\cite{1,16}.

The authors advance three analytical arguments in support of the validity
of the Nernst heat theorem in Ref.~\cite{1}, but no direct proof is given.
As the starting point, two simple systems containing three
harmonic oscillators each are considered. It is shown that for
some ``subsystem'' the entropy may be negative. Needless to say,
systems of this type are not realistic models of full-scale
Quantum Electrodynamical interaction between the plates made of
real metal as they do not reflect the main features of this
interaction. Next, as a ``real metal'' the case where $\e$ is
independent of $\zeta$ is considered. Up to order of magnitude 
estimation for the entropy is obtained consistent with the Nernst
heat theorem. This estimation, however, is not relevant to the
Drude metal because the dielectric permittivity (\ref{3})
depends on frequency. The final analytical argument of
Ref.~\cite{1} is for the frequency-dependent $\e$. In this case,
however, no result is obtained. The authors speculate that
``For a real metal obeying the Drude dispersion relation (3)
(with $\nu\neq 0$),... we expect a $T^3$ (or $T^4$) correction
to the free energy at sufficiently low temperature.'' If this were
the case, the Nernst heat theorem would be satisfied. We show below  
that this is, however, not the case.

To find the asymptotic behavior of the free energy, given by 
Eqs.~(\ref{1})-(\ref{3}), on temperature in the region of low temperatures,
$T\ll T_{eff}=1/(2a)$, one needs the explicit dependence of the
relaxation parameter $\nu$ on temperature. This is given by the
Bloch-Gr\"{u}neisen law through the linear connection between the
relaxation parameter and the static resistivity. As a result,
$\nu\sim T^{\alpha}$ where at low temperatures $\alpha\approx 5$
whereas at high temperatures $\alpha=1$ \cite{1,26,28}
(the tabulated data for different metals can be found in Ref.~\cite{30}).
Note that $\nu(T)\neq 0$ except of one point $T=0$.
Performing the perturbation expansion in Eqs.~(\ref{1})-(\ref{3})
in powers of three small parameters, $T/T_{eff}$, $\nu/\omega_P$ and
$\lambda_p/(2\pi a)$ (we consider separation distances
$a\geq\lambda_p$ where $\lambda_p$ is the plasma wavelength), the following
result for the free energy is obtained
\bes
&&
F=-\frac{\pi^2}{720a^3}\left(1-
2\frac{\lambda_p}{\pi a} +\frac{18}{5}
\frac{\lambda_p^2}{\pi^2 a^2}\right)
-\frac{\zeta(3)}{16\pi a^3}\left[\left(
1+\frac{\lambda_p}{\pi a}\right)\left(\frac{T}{T_{eff}}\right)^3-
\frac{\pi^3}{45\zeta(3)}\left(
1+2\frac{\lambda_p}{\pi a}\right)\left(\frac{T}{T_{eff}}\right)^4
\right]
\nn \\
&&
\phantom{F=}
+\frac{\zeta(3)T}{16\pi a^2}
\left(1-
2\frac{\lambda_p}{\pi a} +3\frac{\lambda_p^2}{\pi^2 a^2}\right)
+\frac{\nu}{\omega_p}\,\frac{T}{4\pi a^2}
\sum\limits_{m=1}^{\infty}\left[{\tilde{\zeta}}_m
\int_{{\tilde{\zeta}}_m}^{\infty}
\frac{dy}{e^y-1}+\frac{1}{{\tilde{\zeta}}_m}
\int_{{\tilde{\zeta}}_m}^{\infty}
\frac{y^2dy}{e^y-1}\right].
\label{4}
\ees
\ni
Here the nondimensional Matsubara frequencies are defined as
${\tilde{\zeta}}_m=2a\zeta_m=4\pi ma/\beta$.
All the details of derivation of perturbation Eq.~(\ref{4})
can be found in Ref.~\cite{26}.

It is quite clear from Eq.~(\ref{4}) that the expectations of 
Ref.~\cite{1} are not borne out . In addition to the higher-order terms in
temperature, there is a linear term which leads to a nonzero 
value of entropy at zero temperature
\beq
S(T=0)=-{\left.\frac{\partial F}{\partial T}\right|}_{T=0}
=-\frac{\zeta(3)}{16\pi a^2}\left(1-
2\frac{\lambda_p}{\pi a} +3\frac{\lambda_p^2}{\pi^2 a^2}\right)<0.
\label{5}
\eeq
\ni
Thus, we confirm the conclusion of Refs.~\cite{26,28} that in
the framework of the Lifshitz formula supplemented by the Drude
model with a temperature-dependent relaxation parameter
(see Fig.~6 of Ref.\cite{1}) the Nernst heat theorem is
in fact violated.

One may mention, as the authors of Ref.~\cite{1} do in the end
of Appendix D, that their dependence of $\nu$ on $T$ neglects
the effect of impurities, which give rise to some nonzero
residual resistivity at zero temperature \cite{31}. This,
however, cannot remedy the situation concerning the inconsistency with 
the Nernst heat theorem. In fact, the resistivity ratio of a sample
can be defined as the ratio of its resistivity at room temperature
to its residual resistivity. For pure samples the
resistivity ratio may be as high as $10^6$ \cite{31}.
As an example let us consider $Au$ with 
$\nu(T=300\,\mbox{K})=5.32\times 10^{13}\,$rad/s \cite{30}.
In this case for the residual value of the relaxation parameter
one obtains $\nu_{res}=5.32\times 10^7\,$rad/s.
The asymptotic expression (\ref{4}) is applicable under the
condition $\nu\ll\zeta_1=2\pi T$ (this condition is automatically
fulfiled for the above used dependence $\nu(T)\sim T^{\alpha}$
with $\alpha\geq 1$). Thus, with allowance made for impurities, the 
asymptotic (\ref{4}) is applicable at temperatures 
$T\gg\nu_{res}/(2\pi)=6.5\times10^{-5}\,$K. What this means is
that the entropy at temperature $T=5\times 10^{-4}\,$K has a nonzero negative
value given by Eq.~(\ref{5}). Physically this is equivalent to
the violation of the Nernst heat theorem. We would like to
point out also that the usual theory of the electron-phonon
interaction, describing the electrons interacting with the
elementary excitations of a perfect lattice with no impurities,
must satisfy and does satisfy all the requirements of
thermodynamics. That is why, the attempt to remedy the violation
of the Nernst heat theorem at the expense of impurities is
meaningless.

We would now like to address the numerical arguments of Ref.~\cite{1}
in support of the validity of the Nernst heat theorem in the
Lifshitz theory combined with the Drude model. All the numerical
computations in Ref.~\cite{1} are performed 
for $Au$ with $\omega_p=9.0\,$eV,
$\nu(T=300\,\mbox{K})=35\,$meV. Once again, no direct computations
by Eqs.~(\ref{1})-(\ref{3}) with $\nu(T)$ given by Fig.~6 of
Ref.~\cite{1} are done. Instead, when calculating
force-temperature relation, the room temperature data for
$\e(i\zeta)$ or the above value of $\nu(T=300\,\mbox{K})$ were
used ignoring the correct
temperature dependence [e.g. our Eq.~(\ref{3}) combined with
Fig.~6 of Ref.~\cite{1} presenting $T$-dependence of $\nu$].
All computed force-temperature and force-distance curves
demonstrate nonmonotonous behavior (see Figs.~2-4 of Ref.~\cite{1}). 
This is a counterintuitive
effect as the authors of Ref.~\cite{1} themselves recognize.
In fact, with an increase of temperature the population of
modes and, consequently, force modulus should increase. 
The nonmonotonous character of the force curves is an artifact.
It is accounted for by the absence of the transverse electric
zero mode contribution for metals in the formalism under
discussion (see below).

In an effort to confirm the validity of the Nernst heat theorem 
in their formalism, the authors of Ref.~\cite{1} compute numerically 
the force-temperature dependence for dielectrics with constant
$\e= 10^2\,\,10^3,\,10^4$, and $\e=\infty$ (Fig.~5 of Ref.~\cite{1}).
They consider the horizontal slope near the point $T=0$, which
is present in all curves, except for $\e=\infty$ and for $Au$
as a desirable property. (``If the force had a linear dependence on
$T$ for small $T$ so would the free energy $F$, in contradiction with
the requirement that the entropy $S=-\partial F/\partial T$ has to go
to zero as $T\to 0$.'') The absence of a desirable slope for $Au$
is considered as a lack of resolution on the scale of the Figure.
It is noted also in \cite{1} that not only dispersive but also
nondispersive curves are nonmonotonous.

All these conclusions depend on the authors' physically
incorrect assumption that there exist dielectrics with arbitrary large 
constant $\e$ and that metals may be considered as a limiting case 
of such kind dielectrics when $\e\to\infty$ on the imaginary
frequency axis. In actual truth,
$\e$ can be assumed to be frequency- and temperature-independent
only in the case of so called non-polar dielectrics whose atoms
or molecules do not have their own dipole moments. The electric
susceptibility of non-polar dielectrics arises due to the electronic
polarization of atoms and molecules. The values of $\e$ for non-polar
dielectrics are of order of one only \cite{30,32,33}.
Large values of $\e$ can exist only for polar dielectrics where
the partial orientation of permanent dipole moments occurs.
But for polar dielectrics $\e$ depends strongly on the frequency and 
temperature. Specifically, their $\e$ quickly decreases with the increase 
of frequency. As a result, at optical and infrared frequencies, which
are characteristic for the Casimir effect, the values of $\e$
are determined by the electronic polarizability \cite{30} and cannot
exceed several units. Thus, the nonmonotonous force-temperature curves
presented in Fig.~5 of Ref.~\cite{1} are not relevant to the resolution
of the above puzzle with the Nernst heat theorem.

The reference to the experiment by Bressi et al \cite{8} where,
supposedly, ``the observed Casimir forces were lower than those
predicted by the traditional (SDM) theory for conducting
plates, in cases where the distances were low, $a\leq 0.5\,\mu$m''
is a misunderstanding. According to Ref.~\cite{1}, this reduction
effect is apparent from Fig.~4 of \cite{8} and could experimentally
confirm the  characteristic temperature variations
predicted by the authors. In Fig.~4 of \cite{8}, however,
all data are presented only for $a>0.5\,\mu$m and there 
is no evidence that these data support expectations of
\cite{1}.

In view of the above, we arrive at the conclusion that the arguments
presented in Ref.~\cite{1} are not correct. The use of the Drude model
in combination with the Lifshitz formula actually leads to the
violation of the Nernst heat theorem and the other nonphysical
features such as a linear temperature correction to the Casimir force
at small separations, negative values of entropy and nonmonotonous
force-temperature and force-distance relations.

Recently the resolution of these complicated problems was obtained
\cite{29} using an alternative approach to the description of
real metals based on the concept of the surface impedance.
This approach offers a fundamental understanding of the reason why
the Drude model is not compatible with the theory of the thermal
Casimir force between real metals. This is due to the fact that the 
Drude dielectric 
function (\ref{3}) is obtained for  the frequency region of the normal
skin effect, where the electromagnetic oscillations penetrate through the
skin layer, and lead to a real current of the conduction
electrons. The interaction of the conduction electrons with the elementary 
excitations of the crystal lattice (phonons) leads to the occurrence
of electric resistance and heating of a metal. By contrast, the thermal
photons in thermal equilibrium with a metal plates or, much less, the
virtual photons giving rise to the Casimir effect cannot, under any
circumstances, lead to the initiation of a real current and heating
of a metal. Of course, this is strictly prohibited by thermodynamics.
Hence the standard concept of a fluctuating electromagnetic field
penetrating inside a metal described by the Drude dielectric function
(\ref{3}) fails to describe virtual and thermal photons. As a consequence,
the Lifshitz formula in combination with the Drude model leads to the
above contradictions with thermodynamics. Note that the dielectric
permittivity depending only on frequency is also inapplicable in the
frequency domain of the anomalous skin effect (see Ref.~\cite{29}
for details).

In contrast to the dielectric permittivity, the surface impedance
$Z(\omega)$ is defined at all frequencies. The impedance boundary
conditions are
\beq
\mbox{\boldmath{$E$}}_t=Z(\omega)\left[
\mbox{\boldmath{$B$}}_t\times\mbox{\boldmath{$n$}}\right],
\label{6}
\eeq
\ni
where $\mbox{\boldmath{$E$}}_t$, $\mbox{\boldmath{$B$}}_t$ are
the tangential components of electric and magnetic fields,
{\boldmath{$n$}} is the internal normal vector to the surface.
They take into account the reflection properties of real metal 
with no consideration of the electromagnetic fluctuations inside of it.

By the use of the surface impedance instead of the Drude model (\ref{3}),
the Lifshitz formula (\ref{1}) is preserved, but the coefficients
$A_m$, $B_m$ from Eq.~(\ref{2}) should be replaced by \cite{29}
\beq
A_m^{imp}=\left(\frac{q-Z_m\zeta_m}{q+Z_m\zeta_m}\right)^2,
\quad
B_m^{imp}=\left(\frac{\zeta_m-Z_mq}{\zeta_m+Z_mq}\right)^2,
\label{7}
\eeq
\ni
where the impedance is computed on the Matsubara frequencies. 
Substituting in Eq.~(\ref{7}) the impedance function of the normal
skin effect or the anomalous skin effect, one finds 
$B_0^{imp}(0)=1$ \cite{29}. In the region of the infrared optics it follows
\cite{29} $B_0^{imp}(0)=(\omega_p-q)^2/(\omega_p+q)^2$.

Let us now present several computational results obtained by Eqs.~(\ref{1}),
(\ref{7}) in the framework of the impedance approach in comparison with the 
results of Ref.~\cite{1}. In Fig.~1, the magnitude of the Casimir 
surface force
density ${\cal{F}}^{T}=-\partial F/\partial T$ for gold is computed,
in the temperature interval 1\,K$\leq T\leq 1200\,$K at a separation
distance $a=1\,\mu$m. Solid line is calculated in the framework of the
impedance approach (separation of $1\,\mu$m corresponds to the domain
of the infrared optics), and dashed line is obtained by the approach 
of Ref.~\cite{1} (i.e. via the Lifshitz formula supplemented by the
Drude model with a temperature dependent relaxation parameter).
It is clearly seen that the dashed line is nonmonotonous demonstrating
the existence of a wide temperature region where force modulus 
decreases with an increase of temperature (like in Figs.~2,\,3 of
Ref.~\cite{1}). At the same time, solid line demonstrates the monotonous
increase of the modulus of the Casimir force with temperature which is
consistent with our expectations on the basis of thermodynamics.

In Fig.~2, the magnitude of the Casimir force density between dielectrics
with $\e=$const is computed at different temperatures at a separation
of $a=1\,\mu$m. Both lines were obtained by the usual Lifshitz formula
(\ref{1}), (\ref{2}) with ${\e}_m=\e=$const. Solid line is for mica
with $\e=7$; dashed line faithfully copies the line of Fig.~5 of
Ref.~\cite{1} with $\e=100$.
Solid line demonstrates the
monotonous increase of the Casimir force with increasing temperature as is
expected from thermodymanics. Dashed line is nonmonotonous because
non-polar dielectrics with so high $\e$ do not exist in nature
due to bounds on the possible values of electronic polarization.
{}From this figure it is clear that the
Lifshitz formula for dielectrics is consistent with
thermodynamics when the correct input data for $\e$ are substituted. 

 In Fig.~3, the Casimir entropy for gold is plotted as a function
of temperature at a separation distance between plates of
$a=1\,\mu$m. Solid line is computed in the framework of the
impedance approach. Dashed line is obtained by the approach 
of Ref.~\cite{1}, i.e. by the usual Lifshitz formula and
Drude model (\ref{1})-(\ref{3}) with a relaxation parameter
depending on temperature according to Fig.~6 of Ref.~\cite{1}.
Evidently, the solid line satisfies all conditions, 
i.e. positive values of entropy at nonzero temperatures, and
the validity of the Nernst heat theorem. By contrast, the dashed
line presents negative values of entropy and the violation
of the Nernst heat theorem (for two semispaces separated by a gap
there is no possibility to introduce some ``composite system''
whose ``subsystem'' with a negative entropy these semispaces
would be; for this reason the Casimir entropy must be positive
which is in fact the case in the impedance approach).
The analytical proof of the validity of the Nernst heat theorem 
in the impedance approach can be found in Ref.~\cite{29}.

Now we are in a position to answer the questions raised in the 
introductory note to this Comment. The answer to the question 
on whether the transverse electric zero mode contributes to the
Casimir effect is yes. As a consequence, there is no linear 
temperature correction to the Casimir force between metallic 
plates at small temperatures. The Casimir free energy is
always decreasing and the force magnitude  increases with
increase of temperature. As to the Casimir entropy, it is
always positive and obeys Nernst's theorem as it must be
in accordance with thermodynamics.
Regarding the doubts not only on the applicability of
the Drude model as such, but even more, doubt on the
applicability of the fundamental Lifshitz formula, they are 
carried too far. Although, as discussed above, the Drude model 
is in fact not appropriate to describe the thermal Casimir effect 
in the case of real metals,
the Lifshitz formula (\ref{1}) with coefficients (\ref{7})
in terms of the surface impedance is perfectly well suited
to calculate all the quantities of physical interest.

\newpage
\widetext
\begin{figure}[h]
\vspace*{-4cm}
\epsfxsize=20cm\centerline{\epsffile{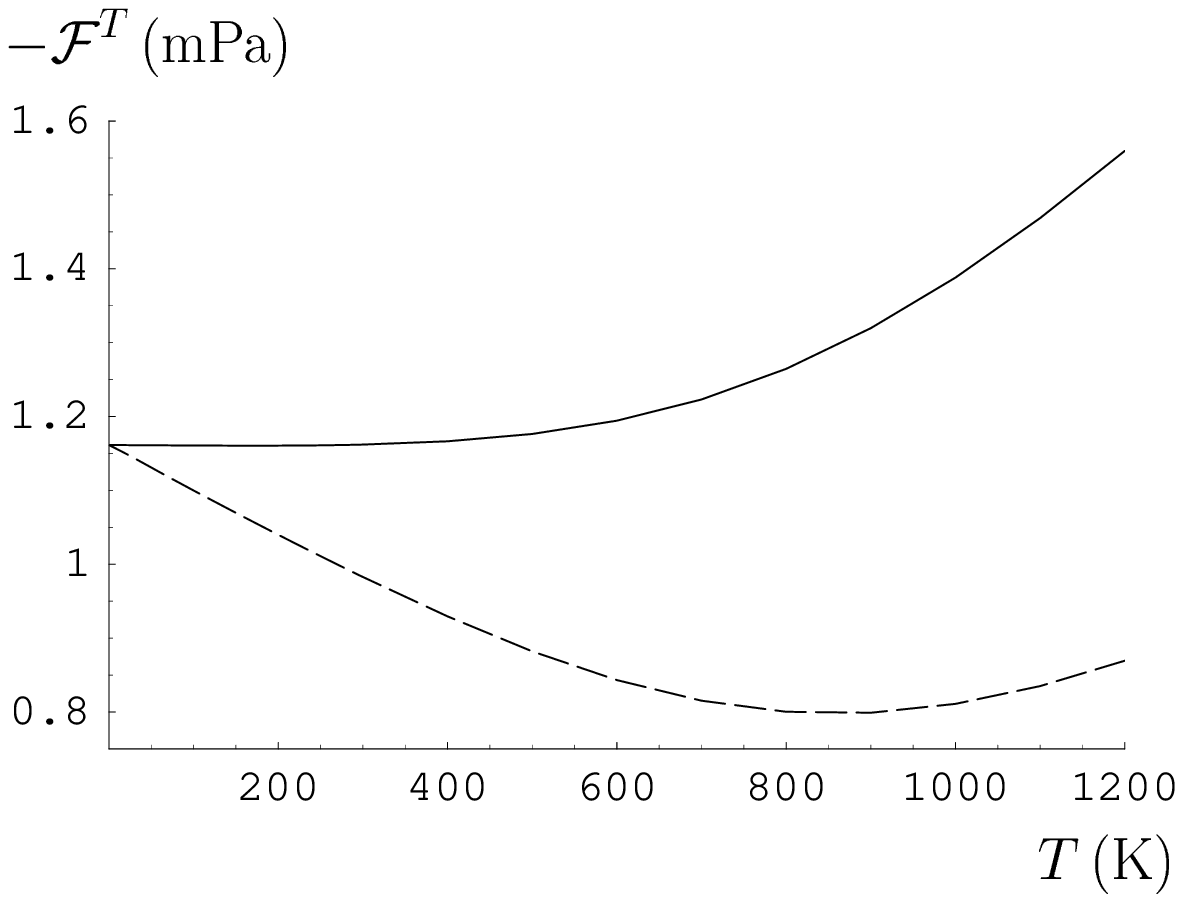}}
\vspace*{-5cm}
\caption{Magnitude of surface force density for gold versus
temperature when $a=1\,\mu$m. The solid line is the physical
result calculated in the framework of the impedance approach.
The dashed line is obtained by the use of the Drude model like
in Ref.~[1].
}
\end{figure}
\newpage
\widetext
\begin{figure}[h]
\vspace*{-7cm}
\epsfxsize=20cm\centerline{\epsffile{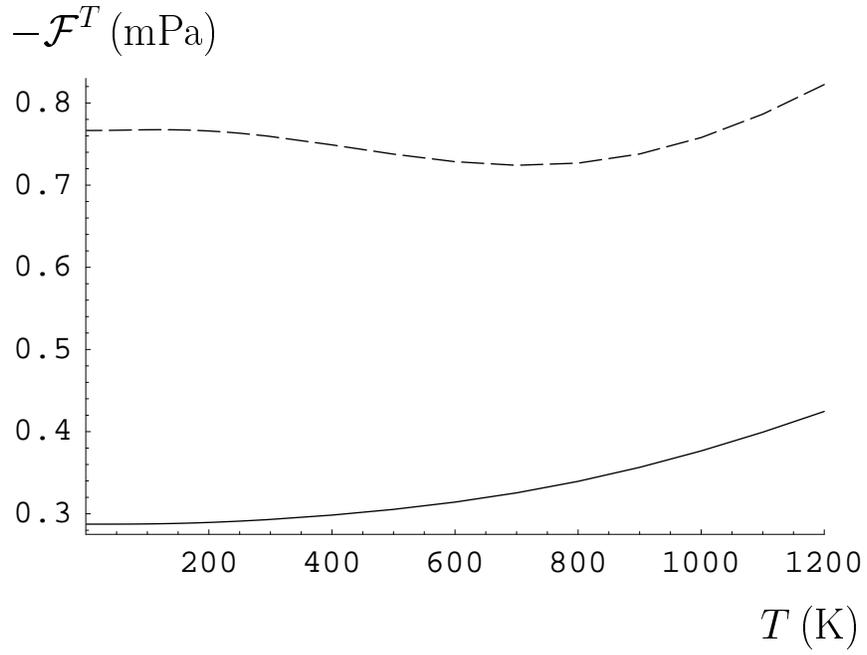}}
\vspace*{-8cm}
\caption{Magnitude of surface force density for 
nondispersive dielectrics versus
temperature when $a=1\,\mu$m. The solid line is the physical
result calculated by the usual Lifshitz formula for mica.
The dashed line is obtained for non-existent non-polar
dielectric with $\e =100$ used in Ref.~[1].
}
\end{figure}
\newpage
\widetext
\begin{figure}[h]
\vspace*{-3cm}
\epsfxsize=20cm\centerline{\epsffile{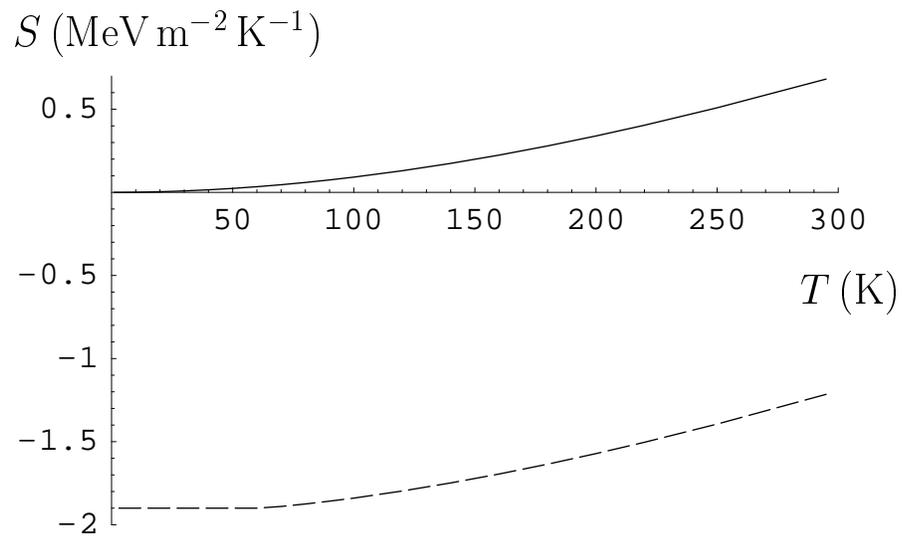}}
\vspace*{-5cm}
\caption{Casimir entropy for two gold semispaces versus temperature
when $a=1\,\mu$m. The solid line is the physical result calculated
in the framework of the impedance approach.
The dashed line is obtained by the use of the Drude model like
in Ref.~[1]. 
}
\end{figure}
\end{document}